\documentclass[a4paper,12pt]{article}

\usepackage{amsmath}
\usepackage{amssymb}
\usepackage{latexsym}
\usepackage{enumerate}
\usepackage{graphicx}
\usepackage{placeins}

\newcommand{\be}{\begin{eqnarray}}
\newcommand{\ee}{\end{eqnarray}}

\newcommand{\bsub}{\begin{subequations}}
\newcommand{\esub}{\end{subequations}}

\newcommand{\disfrac}[1][2]{\displaystyle\frac}
\begin{document}

\title{Energy-momentum for a  charged nonsingular black hole solution with a nonlinear mass function}
\author{I. Radinschi$^{\text{*1}}$, Th. Grammenos$^{\text{**2}}$, Farook Rahaman$%
^{\text{***3}}$, A. Spanou$^{\text{****4}}$,\\
 Sayeedul Islam$^{\text{*****5}}$, Surajit Chattopadhyay$^{\text{******6}}$,\\
  and Antonio Pasqua$^{\text{*******7}}$\\
$^{\text{1}}$Department of Physics \\
``Gh. Asachi'' Technical University, \\
Iasi, 700050, Romania\\
$^{\text{2}}$Department of Civil Engineering, \\
University of Thessaly, 383 34 Volos, Greece\\
$^{\text{3}}$Department of Mathematics, Jadavpur University,\\
Kolkata 700 032, West Bengal, India\\
$^{\text{4}}$School of Applied Mathematics and Physical Sciences,\\
National Technical University of Athens, 157 80 Athens, Greece\\
$^{\text{5}}$Department of Mathematics, Jadavpur University,\\
Kolkata 700 032, West Bengal, India\\
$^{\text{6}}$Department of Mathematics, Amity Institute of Applied Sciences,\\ Amity University, Major Arterial Road, Action Area II,\\ Rajarhat, New Town, West Bengal 700135, India\\
$^{\text{7}}$Department of Physics, University of Trieste, Via Valerio, 2,\\ 34127 Trieste, Italy\\
$^{\text{*}}$radinschi@yahoo.com, $^{\text{**}}$thgramme@civ.uth.gr,\\
$^{\text{***}}$rahaman@iucaa.ernet.in, $^{\text{****}}$aspanou@central.ntua.gr,%
\\
$^{\text{*****}}$sayeedul.jumath@gmail.com,\\ 
$^{\text{******}}$surajcha@associates.iucaa.in, $^{\text{*******}}$toto.pasqua@gmail.com}
\date{}
\maketitle

\begin{abstract}
The energy-momentum of a new four-dimensional, charged, spherically symmetric and nonsingular  black hole solution constructed in the context of general relativity coupled to a theory of nonlinear electrodynamics  is investigated, whereby the nonlinear mass function is inspired by the probability density function of the continuous logistic distribution. The energy and momentum distributions are calculated by use of the Einstein, Landau-Lifshitz, Weinberg and M\o ller energy-momentum complexes. In all these prescriptions it is found that the energy distribution depends on the mass $M$ and the charge $q$ of the black hole, an additional parameter $\beta$ coming from the gravitational background considered, and on the radial coordinate $r$. Further, the Landau-Lifshitz and Weinberg prescriptions yield the same result for the energy, while in all the aforesaid prescriptions all the momenta vanish. We also focus on the study of the limiting behavior of the energy for different va!
 lues of the radial coordinate, the parameter $\beta$, and the charge $q$. Finally, it is pointed out that for $r\rightarrow \infty$ and $q = 0$ all the energy-momentum complexes yield the same expression for the energy distribution as in the case of the Schwarzschild black hole solution.\\

\end{abstract}

\section{Introduction}

The problem  of the energy-momentum localization in General Relativity  has been investigated  over the years by using various and different powerful tools such as superenergy tensors \cite{Bel}-\cite{Senovilla}, quasi-local expressions \cite{Brown}-\cite{Balart_1}, and the
mostly known pseudo-tensorial energy-momentum complexes introduced by Einstein \cite{Einstein},\cite{Trautman}, Landau and Lifshitz \cite{Landau}, Papapetrou \cite{Papapetrou}, Bergmann and Thomson \cite{Bergmann}, M\o ller \cite{Moller}, Weinberg \cite{Weinberg}, and
Qadir and Sharif \cite{Qadir}. 

As it is well-known, the main difficulty which arises consists in developing a properly defined expression for the energy density of the gravitational background. Until today, no generally accepted meaningful definition for the energy of the gravitational field has been established. However, despite this difficulty, many physically reasonable results have been obtained by applying  the aforesaid definitions for the energy-momentum localization. At this point, one cannot but notice the existing agreement between the pseudo-tensorial prescriptions and the quasi-local mass definition elaborated by Penrose \cite{Penrose} and further developed by Tod \cite{Tod}.

Although the dependence on the coordinate system continues to be the main ``weakness'' of these tools, a number of physically interesting results have been obtained for gravitating systems in  $(3+1)$, $(2+1)$, and $(1+1)$ spacetime dimensions by using the energy-momentum complexes \cite{Virbhadra}-\cite{Worm}. In fact, the M\o ller energy-momentum complex is the only computational tool independent of coordinates. In the context of other pseudo-tensorial prescriptions, in order to calculate the energy and momentum distributions one introduces Schwarzschild  Cartesian and  Kerr-Schild  coordinates. 

An alternative option for avoiding the problem of the coordinate system dependence is provided by the  teleparallel theory of gravity \cite{Rag}-\cite{Aldrovandi}, whereby one notices the considerable similarity of results obtained by this approach with results achieved by using the energy-momentum complexes \cite{Gamal}-\cite{Jawad}. 

Finally, closing this short introduction to the topic of the energy-momentum localization, it is necessary to point out the broadness of the ongoing attempts in order to define properly and, actually, rehabilitate the concept of the energy-momentum complex \cite{Chang}-\cite{Sun_Nester}.

The outline of the present paper is the following. In Section 2 we introduce  the new static and charged, spherically symmetric, nonsingular  black hole solution under study. Section 3 is devoted to the presentations of the Einstein, Landau-Lifshitz, Weinberg, and M\o ller energy-momentum complexes used for the calculations. In Section 4 the computations of the energy and momentum distributions are presented. In the Discussion given in Section 5, we comment on our results and explore  some limiting and particular cases. We have used  geometrized units $(c=G=1)$, while the signature  is $(+,-,-,-)$. The calculations for the Einstein, Landau-Lifshitz, and Weinberg
energy-momentum complexes are performed by use of the Schwarzschild Cartesian
coordinates. Greek indices range from $0$ to $3$, while Latin indices run
from $1$ to $3$.

\section{The New Charged Nonsingular Black Hole Solution with a nonlinear mass function}

The determination of nonsingular black hole solutions by coupling gravity to nonlinear electrodynamics has attracted interest long ago (for a review see, e.g., \cite{Ansoldi} for spherically symmetric solutions, or \cite{Dymnikova}  and references therein for charged axisymmetric solutions).
Recently,  L. Balart and E. C. Vagenas \cite{Balart_3} constructed a number of new charged, nonsingular and spherically symmetric, four-dimensional black hole solutions with a nonlinear electrodynamics
source. Indeed, starting with the static and spherically symmetric spacetime geometry described by the line element
\begin{equation}\label{line_element}
ds^2=-f(r)dt^2+\frac{1}{f(r)}dr^2+r^2(d\theta^2+\sin^2\theta d\phi^2),
\end{equation}
with the metric function
\begin{equation}\label{metric_function}
f(r)=1-\frac{2M}{r}\left[\frac{\sigma(\beta r)}{\sigma_\infty}\right]^\beta,
\end{equation}
where the distribution function $\sigma(\beta r)$ depends on the mass $M$, the charge $q$, the radial coordinate $r$, and the parameter $\beta \in \mathbb{R}^+$, and $\sigma_\infty=\sigma (r\rightarrow \infty )$ is a normalization factor. Thus one has a nonlinear mass function of the form
\begin{equation}\label{mass_function}
m(r)=M\left[\frac{\sigma(\beta r)}{\sigma_\infty}\right]^\beta,
\end{equation}
which, at infinity, becomes $M$. It is shown in \cite{Balart_3} that for specific distribution functions $\sigma(r)$ the curvature invariants ($R$, $R_{\mu\nu}R^{\mu\nu}$, $R_{\kappa\lambda\mu\nu}R^{\kappa\lambda\mu\nu}$) and the associated nonlinear electric field are nonsingular everywhere.

Based on these results, we have already studied \cite{Radinschi_AHEP} the problem of the localization of energy for a function $\sigma(r)$ resembling the form of the Fermi-Dirac distribution. Here, following up that work, we adopt another distribution function given in \cite{Balart_3} that is inspired by the form of the probability density function of the continuous logistic distribution \cite{Balakrishnan}, such that 
\begin{equation}\label{THE_metric_function}
f(r)=1-\frac{2M}{r}\left\{\frac{4 \exp \left(-\sqrt{\frac{2q^{2}}{\beta Mr}}\right)}{\left[1+\exp \left(-\sqrt{\frac{2q^{2}}{\beta Mr}}\right)\right]^{2}}\right\}^{\beta}, \qquad \beta\in \mathbb{R}^+
\end{equation}
with the nonlinear mass function given by (\ref{mass_function}).

The associated nonsingular spacetime exhibits two horizons, while the nonlinear and nonsingular electric field that asymptotically goes to $q/r^2$ reads now
\begin{align}\label{electric_field}
E(r)= &\frac{q}{8r^2}\left(\text{sech} \sqrt{\frac{q^2}{2\beta M r}}\right)^{2(1+\beta)}\nonumber\\
&\times \left[(1+\beta)-\beta\cosh \sqrt{\frac{2q^2}{\beta M r}}+7\sqrt{\frac{\beta Mr}{2q^2}}\sinh\sqrt{\frac{2q^2}{\beta M r}}\right].
\end{align}
When $\beta \rightarrow 0$, the metric function (\ref{THE_metric_function}) takes the form
$f(r)=1-2M/r$, i.e. we get the Schwarzschild black hole geometry.

Thus, in what follows we are going to investigate the problem of energy-momentum localization for a charged and nonsingular black hole solution with the spacetime described by (\ref{line_element}), (\ref{THE_metric_function}) and the nonlinear mass function 
\begin{equation}\label{nonlinear_mass_function}
m(r)=M\left\{\frac{4 \exp \left(-\sqrt{\frac{2q^{2}}{\beta Mr}}\right)}{\left[1+\exp \left(-\sqrt{\frac{2q^{2}}{\beta Mr}}\right)\right]^{2}}\right\}^{\beta}.
\end{equation}

\section{Einstein, Landau-Lifshitz, Weinberg and M\o ller Energy-Momentum
Complexes}

The definition of the Einstein energy-momentum complex \cite{Einstein},\cite{Trautman} for a $(3+1)$
dimensional gravitating system is given by 
\begin{equation}
\theta _{\nu }^{\mu }=\frac{1}{16\pi }h_{\nu ,\,\lambda }^{\mu \lambda }, 
\end{equation}%
where the von Freud superpotentials $h_{\nu }^{\mu \lambda }$ are
given as 
\begin{equation}
h_{\nu }^{\mu \lambda }=\frac{1}{\sqrt{-g}}g_{\nu \sigma }[-g(g^{\mu \sigma
}g^{\lambda \kappa }-g^{\lambda \sigma }g^{\mu \kappa })]_{,\kappa } 
\end{equation}%
and satisfy the required antisymmetric property 
\begin{equation}
h_{\nu }^{\mu \lambda }=-h_{\nu }^{\lambda \mu }.  
\end{equation}%
The components $\theta _{0}^{0}$ and $\theta _{i}^{0}$ correspond to the
energy and the momentum densities, respectively. In the Einstein
prescription the local conservation law holds: 
\begin{equation}
\theta _{\nu ,\mu }^{\mu }=0.  
\end{equation}%
Thus, the energy and the momenta can be computed by 
\begin{equation}
P_{\nu }=\iiint \theta _{\nu }^{0}\,dx^{1}dx^{2}dx^{3}. 
\end{equation}%
Applying Gauss' theorem, the energy-momentum is 
\begin{equation}\label{Einstein_energy}
P_{\nu }=\frac{1}{16\pi }\iint h_{\nu }^{0i}n_{i}dS, 
\end{equation}%
where $n_{i}$ represents the outward unit normal vector on the surface $dS$.

The Landau-Lifshitz energy-momentum complex \cite{Landau} is defined as 
\begin{equation}
L^{\mu \nu }=\frac{1}{16\pi }S_{,\,\rho \sigma }^{\mu \nu \rho \sigma }, 
\end{equation}%
where the Landau-Lifshitz superpotentials are given by: 
\begin{equation}
S^{\mu \nu \rho \sigma }=-g(g^{\mu \nu }g^{\rho \sigma }-g^{\mu \rho }g^{\nu
\sigma }).  
\end{equation}%
The $L^{00}$ and $L^{0i}$ components represent the energy
and the momentum densities, respectively. In the Landau-Lifshitz
prescription the local conservation law reads: 
\begin{equation}
L_{,\,\nu }^{\mu \nu }=0. 
\end{equation}%
By integrating $L^{\mu \nu }$ over the 3-space one obtains for the
energy-momentum: 
\begin{equation}
P^{\mu }=\iiint L^{\mu 0}\,dx^{1}dx^{2}dx^{3}. 
\end{equation}%
By using Gauss' theorem we have 
\begin{equation}\label{Landau_energy}
P^{\mu }=\frac{1}{16\pi }\iint S_{,\nu }^{\mu 0i\nu }n_{i}dS=\frac{1}{%
16\pi }\iint U^{\mu 0i}n_{i}dS. 
\end{equation}

The Weinberg energy-momentum complex \cite{Weinberg} is given by the expression 
\begin{equation}
W^{\mu \nu }=\frac{1}{16\pi }D_{,\,\lambda }^{\lambda \mu \nu }, 
\end{equation}%
where $D^{\lambda \mu \nu }$ are the corresponding superpotentials: 
\begin{equation}
D^{\lambda \mu \nu }=\frac{\partial h_{\kappa }^{\kappa }}{\partial
x_{\lambda }}\eta ^{\mu \nu }-\frac{\partial h_{\kappa }^{\kappa }}{\partial
x_{\mu }}\eta ^{\lambda \nu }-\frac{\partial h^{\kappa \lambda }}{\partial
x^{\kappa }}\eta ^{\mu \nu }+\frac{\partial h^{\kappa \mu }}{\partial
x^{\kappa }}\eta ^{\lambda \nu }+\frac{\partial h^{\lambda \nu }}{\partial
x_{\mu }}-\frac{\partial h^{\mu \nu }}{\partial x_{\lambda }}, 
\end{equation}%
and 
\begin{equation}
h_{\mu \nu }=g_{\mu \nu }-\eta _{\mu \nu }. 
\end{equation}%
Here the $W^{00}$ and $W^{0i}$ components correspond to the energy and the
momentum densities, respectively. In the Weinberg prescription the local
conservation law reads: 
\begin{equation}
W_{,\,\nu }^{\mu \nu }=0. 
\end{equation}%
The integration of $W^{\mu \nu }$ over the 3-space yields for the
energy-momentum: 
\begin{equation}
P^{\mu }=\iiint W^{\mu 0}\,dx^{1}dx^{2}dx^{3}.
\end{equation}%
Applying Gauss' theorem and integrating over the surface of a sphere of
radius $r$, one obtains for the energy-momentum distribution the expression: 
\begin{equation}\label{Weinberg_energy}
P^{\mu }=\frac{1}{16\pi }\iint D^{i0\mu }n_{i}dS.
\end{equation}

The M{\o }ller energy-momentum complex \cite{Moller} is given by the expression 
\begin{equation}
\mathcal{J}_{\nu }^{\mu }=\frac{1}{8\pi }M_{\nu \,\,,\,\lambda }^{\mu
\lambda }, 
\end{equation}%
where the M{\o }ller superpotentials $M_{\nu }^{\mu \lambda }$ are 
\begin{equation}
M_{\nu }^{\mu \lambda }=\sqrt{-g}\left( \frac{\partial g_{\nu \sigma }}{%
\partial x^{\kappa }}-\frac{\partial g_{\nu \kappa }}{\partial x^{\sigma }}%
\right) g^{\mu \kappa }g^{\lambda \sigma }
\end{equation}%
and satisfy the necessary antisymmetric property: 
\begin{equation}
M_{\nu }^{\mu \lambda }=-M_{\nu }^{\lambda \mu }. 
\end{equation}%
M{\o }ller's energy-momentum complex also satisfies the local conservation
law 
\begin{equation}
\frac{\partial \mathcal{J}_{\nu }^{\mu }}{\partial x^{\mu }}=0, 
\end{equation}%
with $\mathcal{J}_{0}^{0}$ and $\mathcal{J}_{i}^{0}$ representing the
energy and the momentum densities, respectively. In the M{\o }ller
prescription, the energy-momentum is given by 
\begin{equation}
P_{\nu }=\iiint \mathcal{J}_{\nu }^{0}dx^{1}dx^{2}dx^{3}. 
\end{equation}%
With the aid of Gauss' theorem one gets%
\begin{equation}\label{Moller_energy}
P_{\nu }=\frac{1}{8\pi }\iint M_{\nu }^{0i}n_{i}dS. 
\end{equation}

\section{Energy and Momentum Distributions for the New Charged Nonsingular Black
Hole Solution}

In order to calculate the energy and momenta by using the Einstein
energy-momentum complex, it is required to transform the metric given by the
line element (\ref{line_element}) in Schwarzschild Cartesian coordinates. We obtain the line
element in the following form: 
\begin{equation}\label{line_element_nonsingular}
ds^{2}=f(r)dt^{2}-(dx^{2}+dy^{2}+dz^{2})-\frac{f^{-1}(r)-1}{r^{2}}%
(xdx+ydy+zdz)^{2}\text{.} 
\end{equation}

The components of the superpotential $h_{\nu }^{0i}$ in Schwarzschild
Cartesian coordinates  are
\begin{equation}\label{Einstein_components}
\begin{split}
h_{1}^{01} &=h_{1}^{02}=h_{1}^{03}=0,\\
h_{2}^{01} &=h_{2}^{02}=h_{2}^{03}=0,\\
h_{3}^{01}&=h_{3}^{02}=h_{3}^{03}=0.
\end{split}
\end{equation}

The remaining, non-vanishing, components of the
superpotentials in the Einstein prescription are: 
\begin{equation}\label{Einstein_super_1}
h_{0}^{01}=\frac{2x}{r^{2}}\frac{2M}{r}\left\{\frac{4 \exp \left(-\sqrt{\frac{2q^{2}}{\beta Mr}%
}\right)}{\left[1+\exp \left(-\sqrt{\frac{2q^{2}}{\beta Mr}}\right)\right]^{2}}\right\}^{\beta },
\end{equation}%
\begin{equation}\label{Einstein_super_2}
h_{0}^{02}=\frac{2y}{r^{2}}\frac{2M}{r}\left\{\frac{4 \exp \left(-\sqrt{\frac{2q^{2}}{\beta Mr}%
}\right)}{\left[1+\exp \left(-\sqrt{\frac{2q^{2}}{\beta Mr}}\right)\right]^{2}}\right\}^{\beta },
\end{equation}%
\begin{equation}\label{Einstein_super_3}
h_{0}^{03}=\frac{2z}{r^{2}}\frac{2M}{r}\left\{\frac{4 \exp \left(-\sqrt{\frac{2q^{2}}{\beta Mr}%
}\right)}{\left[1+\exp \left(-\sqrt{\frac{2q^{2}}{\beta Mr}}\right)\right]^{2}}\right\}^{\beta }.
\end{equation}%
From the line element (\ref{line_element_nonsingular}), the expression 
 (\ref{Einstein_energy}) and the superpotentials (\ref{Einstein_super_1})-(\ref{Einstein_super_3}), we
obtain for the energy distribution in the Einstein prescription (see Figure 1): 
\begin{equation}\label{Einstein_pre}
E_{E}=M\left\{\frac{4 \exp \left(-\sqrt{\frac{2q^{2}}{\beta Mr}%
}\right)}{\left[1+\exp \left(-\sqrt{\frac{2q^{2}}{\beta Mr}}\right)\right]^{2}}\right\}^{\beta }.
\end{equation}
By using (\ref{Einstein_energy}) and (\ref{Einstein_components}) we find
that all the momenta vanish:
\begin{equation}
P_{x}=P_{y}=P_{z}=0. 
\end{equation}

\begin{figure}[t!]
\begin{center}
\includegraphics[width=84mm]{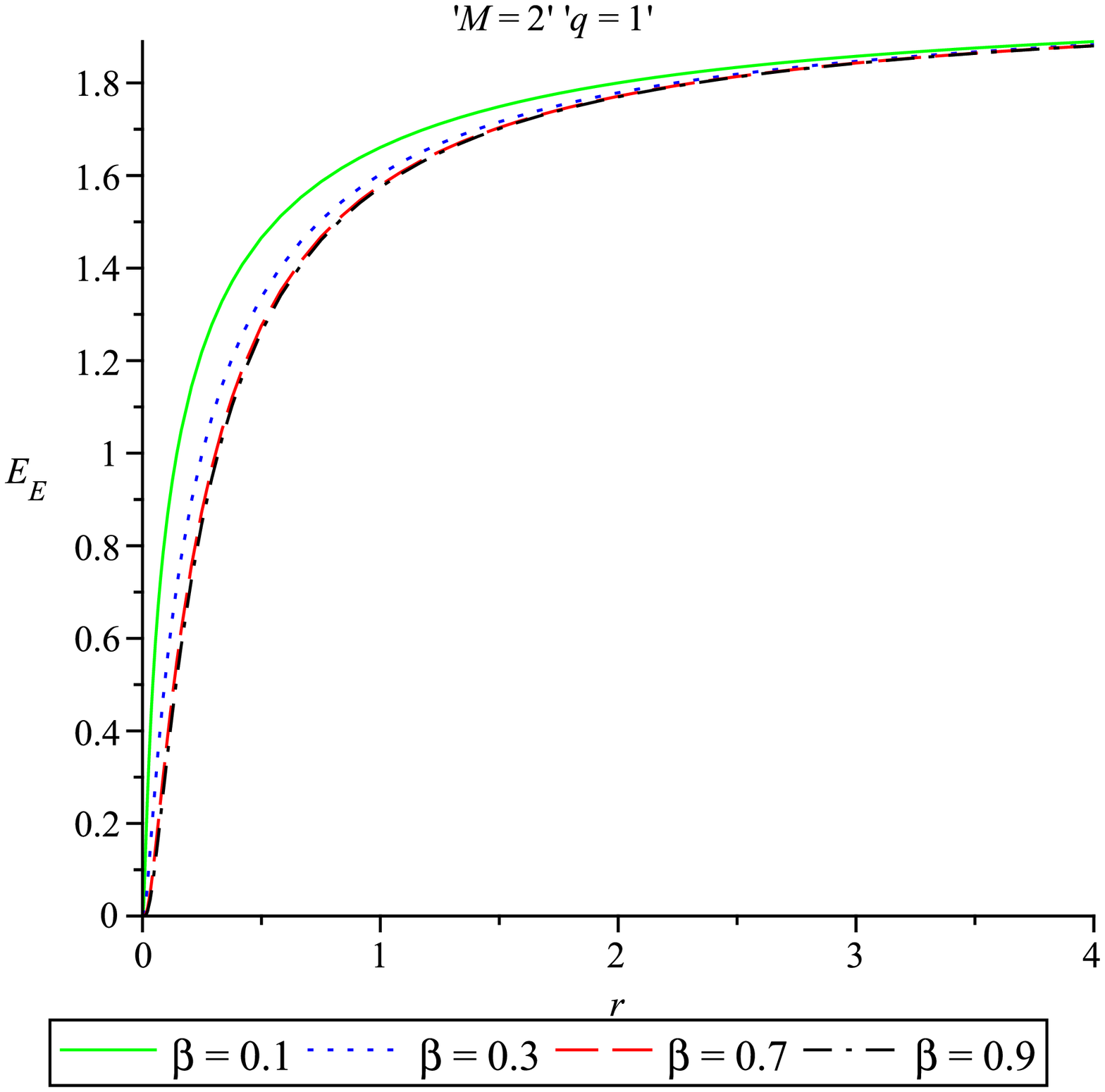}
\caption{Energy distribution obtained in the Einstein prescription for different values of $\beta$.}
\label{fig1}
\end{center}
\end{figure}

In order to apply the Landau-Lifshitz energy-momentum complex, we use the $%
U^{\mu 0i}$ quantities defined in (\ref{Landau_energy}) and we find the
following nonvanishing components: 
\begin{equation}\label{Landau_super_1}
U^{001}=\frac{2x}{r^{2}}
\frac{\displaystyle{\frac{2M}{r}}\left\{\frac{4 \exp \left(-\sqrt{\frac{2q^{2}}{\beta Mr}%
}\right)}{\left[1+\exp \left(-\sqrt{\frac{2q^{2}}{\beta Mr}}\right)\right]^{2}}\right\}^{\beta }}
{1-\displaystyle{\frac{2M}{r}}\left\{\frac{4 \exp \left(-\sqrt{\frac{2q^{2}}{\beta Mr}%
}\right)}{\left[1+\exp \left(-\sqrt{\frac{2q^{2}}{\beta Mr}}\right)\right]^{2}}\right\}^{\beta }},
\end{equation}%
\begin{equation}\label{Landau_super_2}
U^{002}=\frac{2y}{r^{2}}
\frac{\displaystyle{\frac{2M}{r}}\left\{\frac{4 \exp \left(-\sqrt{\frac{2q^{2}}{\beta Mr}%
}\right)}{\left[1+\exp \left(-\sqrt{\frac{2q^{2}}{\beta Mr}}\right)\right]^{2}}\right\}^{\beta }}
{1-\displaystyle{\frac{2M}{r}}\left\{\frac{4 \exp \left(-\sqrt{\frac{2q^{2}}{\beta Mr}%
}\right)}{\left[1+\exp \left(-\sqrt{\frac{2q^{2}}{\beta Mr}}\right)\right]^{2}}\right\}^{\beta }},  
\end{equation}%
\begin{equation}\label{Landau_super_3}
U^{003}=\frac{2z}{r^{2}}
\frac{\displaystyle{\frac{2M}{r}}\left\{\frac{4 \exp \left(-\sqrt{\frac{2q^{2}}{\beta Mr}%
}\right)}{\left[1+\exp \left(-\sqrt{\frac{2q^{2}}{\beta Mr}}\right)\right]^{2}}\right\}^{\beta }}
{1-\displaystyle{\frac{2M}{r}}\left\{\frac{4 \exp \left(-\sqrt{\frac{2q^{2}}{\beta Mr}%
}\right)}{\left[1+\exp \left(-\sqrt{\frac{2q^{2}}{\beta Mr}}\right)\right]^{2}}\right\}^{\beta }}.
\end{equation}%
Now, replacing (\ref{Landau_super_1})-(\ref{Landau_super_3}) in (\ref{Landau_energy}), we obtain the energy distribution in the Landau-Lifshitz prescription: 
\begin{equation}\label{Landau_pre}
E_{LL}=\frac{M\left\{\frac{4 \exp \left(-\sqrt{\frac{2q^{2}}{\beta Mr}%
}\right)}{\left[1+\exp \left(-\sqrt{\frac{2q^{2}}{\beta Mr}}\right)\right]^{2}}\right\}^{\beta }} 
{1-\displaystyle{\frac{2M}{r}}\left\{\frac{4 \exp \left(-\sqrt{\frac{2q^{2}}{\beta Mr}%
}\right)}{\left[1+\exp \left(-\sqrt{\frac{2q^{2}}{\beta Mr}}\right)\right]^{2}}\right\}^{\beta }}, 
\end{equation}%
while all the momenta vanish
\begin{equation}
P^{x}=P^{y}=P^{z}=0.  
\end{equation}

In the Weinberg prescription the  non-vanishing superpotential components are: 
\begin{equation}\label{Weinberg_super_1}
D^{100}=\frac{2x}{r^{2}}
\frac{\displaystyle{\frac{2M}{r}}\left\{\frac{4 \exp \left(-\sqrt{\frac{2q^{2}}{\beta Mr}%
}\right)}{\left[1+\exp \left(-\sqrt{\frac{2q^{2}}{\beta Mr}}\right)\right]^{2}}\right\}^{\beta }}
{1-\displaystyle{\frac{2M}{r}}\left\{\frac{4 \exp \left(-\sqrt{\frac{2q^{2}}{\beta Mr}%
}\right)}{\left[1+\exp \left(-\sqrt{\frac{2q^{2}}{\beta Mr}}\right)\right]^{2}}\right\}^{\beta }},
\end{equation}%
\begin{equation}\label{Weinberg_super_2}
D^{200}=\frac{2y}{r^{2}}
\frac{\displaystyle{\frac{2M}{r}}\left\{\frac{4 \exp \left(-\sqrt{\frac{2q^{2}}{\beta Mr}%
}\right)}{\left[1+\exp \left(-\sqrt{\frac{2q^{2}}{\beta Mr}}\right)\right]^{2}}\right\}^{\beta }}
{1-\displaystyle{\frac{2M}{r}}\left\{\frac{4 \exp \left(-\sqrt{\frac{2q^{2}}{\beta Mr}%
}\right)}{\left[1+\exp \left(-\sqrt{\frac{2q^{2}}{\beta Mr}}\right)\right]^{2}}\right\}^{\beta }}, 
\end{equation}%
\begin{equation}\label{Weinberg_super_3}
D^{300}=\frac{2z}{r^{2}}
\frac{\displaystyle{\frac{2M}{r}}\left\{\frac{4 \exp \left(-\sqrt{\frac{2q^{2}}{\beta Mr}%
}\right)}{\left[1+\exp \left(-\sqrt{\frac{2q^{2}}{\beta Mr}}\right)\right]^{2}}\right\}^{\beta }}
{1-\displaystyle{\frac{2M}{r}}\left\{\frac{4 \exp \left(-\sqrt{\frac{2q^{2}}{\beta Mr}%
}\right)}{\left[1+\exp \left(-\sqrt{\frac{2q^{2}}{\beta Mr}}\right)\right]^{2}}\right\}^{\beta }}. 
\end{equation}%
Substituting the expressions obtained in (\ref{Weinberg_super_1})-(\ref{Weinberg_super_3}) into (\ref{Weinberg_energy}) we get for the
energy distribution inside a $2$-sphere of radius $r$ the expression: 
\begin{equation}\label{Weinberg_pre}
E_{W}=\frac{M\left\{\displaystyle{\frac{4 \exp \left(-\sqrt{\frac{2q^{2}}{\beta Mr}%
}\right)}{\left[1+\exp \left(-\sqrt{\frac{2q^{2}}{\beta Mr}}\right)\right]^{2}}}\right\}^{\beta }} 
{1-\displaystyle{\frac{2M}{r}}\left\{\frac{4 \exp \left(-\sqrt{\frac{2q^{2}}{\beta Mr}%
}\right)}{\left[1+\exp \left(-\sqrt{\frac{2q^{2}}{\beta Mr}}\right)\right]^{2}}\right\}^{\beta }}. 
\end{equation}%
In the Weinberg prescription all the momenta vanish
\begin{equation}
P^{x}=P^{y}=P^{z}=0. 
\end{equation}

One can see from  (\ref{Landau_pre}) and (\ref{Weinberg_pre}) that the energy in the Landau-Lifshitz
prescription is identical with the energy in the Weinberg prescription.

Finally, in the M\o ller prescription, we find only one non-vanishing superpotential: 
\begin{equation}\label{Moller_super}
M_{0}^{01}=2M\sin \theta 
\left[\displaystyle{\frac{4\exp \left(-\sqrt{\frac{2q^{2}}{\beta Mr}
}\right)}{\left[1+\exp \left(-\sqrt{\frac{2q^{2}}{\beta Mr}}\right)\right]^{2}}}\right]^{\beta }
\left\{
1-\frac{q^2}{Mr^3\sqrt{\frac{2q^2}{\beta Mr}}}
\left[\frac{1-\exp\left(-\sqrt{\frac{2q^2}{\beta Mr}}\right)}{1+\exp\left(-\sqrt{\frac{2q^2}{\beta Mr}}\right)}\right]
\right\}.
\end{equation}

 Using the above expression for the superpotential and with the aid of
the metric coefficient (\ref{THE_metric_function}) and the expression for energy given by (\ref{Moller_energy}), we
obtain the energy distribution in the M\o ller prescription (see Figure 2): 
\begin{equation}
E_{M}=M
\left[\displaystyle{\frac{4 \exp \left(-\sqrt{\frac{2q^{2}}{\beta Mr}
}\right)}{\left[1+\exp \left(-\sqrt{\frac{2q^{2}}{\beta Mr}}\right)\right]^{2}}}\right]^{\beta }
\left\{
1-\sqrt{\frac{q^2\beta}{2Mr}}
\left[\frac{1-\exp\left(-\sqrt{\frac{2q^2}{\beta Mr}}\right)}{1+\exp\left(-\sqrt{\frac{2q^2}{\beta Mr}}\right)}\right]
\right\},
\end{equation}
while all the momenta vanish
\begin{equation}
P_{r}=P_{\theta }=P_{\phi}=0. 
\end{equation}

\begin{figure}[t!]
\begin{center}
\includegraphics[width=84mm]{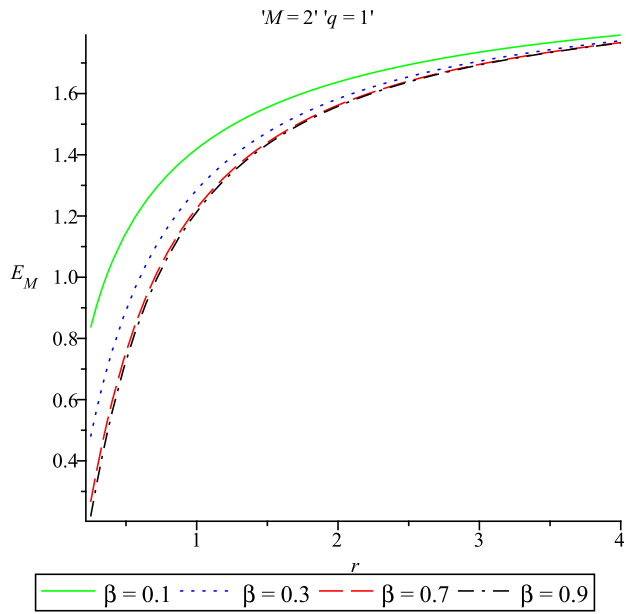}
\caption{Energy distribution obtained in the M\o ller prescription for different values of $\beta$.}
\label{fig1}
\end{center}
\end{figure}

\section{Discussion}
Following up our previous results \cite{Radinschi_AHEP}, we have studied the problem of energy-momentum localization for a new charged, nonsingular and spherically symmetric, static black hole solution in (3+1) dimensions with a nonlinear mass function recently constructed by L. Balart and E.C. Vagenas \cite{Balart_3}. To this purpose, we have applied the Einstein, Landau-Lifshitz, Weinberg, and  M\o ller  pseudotensorial prescriptions. The calculations have shown that, in all the four prescriptions, the momenta vanish, while the energy depends (see, Table 1) on the black hole's mass and charge, the radial coordinate, and a parameter $\beta \in \mathbb{R}^+$ that is a scaling factor of $r$ (in essence a dilation factor, as it is always positive) inspired by
the form of the logistic distribution and marks the spacetime geometry considered. In fact, for each value of $\beta$ one can
numerically determine the values of the two existing horizon radii (an inner and an outer)
obtained from the metric function. It is pointed out that the Landau-Lifshitz and the Weinberg prescriptions yield the same energy distribution.

\begin{table}[h!]
\centering
\begin{tabular}{|c|c|}
\hline
\textbf{Prescription} & \textbf{Energy} \\
\hline 
&\\
Einstein  & $E_{E}=M\left\{\displaystyle{\frac{4\exp \left(-\sqrt{\frac{2q^{2}}{\beta Mr}%
}\right)}{\left[1+\exp \left(-\sqrt{\frac{2q^{2}}{\beta Mr}}\right)\right]^{2}}}\right\}^{\beta }$\\
&\\
\hline
&\\
Landau-Lifshitz & $E_{LL}=\displaystyle{\frac{M\left\{\displaystyle{\frac{4\exp \left(-\sqrt{\frac{2q^{2}}{\beta Mr}
}\right)}{\left[1+\exp \left(-\sqrt{\frac{2q^{2}}{\beta Mr}}\right)\right]^{2}}}\right\}^{\beta }}
{1-\displaystyle{\frac{2M}{r}}\left\{\frac{4\exp \left(-\sqrt{\frac{2q^{2}}{\beta Mr}
}\right)}{\left[1+\exp \left(-\sqrt{\frac{2q^{2}}{\beta Mr}}\right)\right]^{2}}\right\}^{\beta}}}$\\
&\\
\hline
&\\
Weinberg & $E_{W}=\displaystyle{\frac{M\left\{\displaystyle{\frac{4\exp \left(-\sqrt{\frac{2q^{2}}{\beta Mr}
}\right)}{\left[1+\exp \left(-\sqrt{\frac{2q^{2}}{\beta Mr}}\right)\right]^{2}}}\right\}^{\beta }}
{1-\displaystyle{\frac{2M}{r}}\left\{\frac{4\exp \left(-\sqrt{\frac{2q^{2}}{\beta Mr}
}\right)}{\left[1+\exp \left(-\sqrt{\frac{2q^{2}}{\beta Mr}}\right)\right]^{2}}\right\}^{\beta}}}.$\\
&\\

\hline
&\\
  & $E_{M}=M
\left[\displaystyle{\frac{4\exp \left(-\sqrt{\frac{2q^{2}}{\beta Mr}
}\right)}{\left[1+\exp \left(-\sqrt{\frac{2q^{2}}{\beta Mr}}\right)\right]^{2}}}\right]^{\beta }\times$\\
M\o ller &\\
&
$\left\{
1-\sqrt{\displaystyle{\frac{q^2\beta}{2Mr}}}
\left[\displaystyle{\frac{1-\exp\left(-\sqrt{\frac{2q^2}{\beta Mr}}\right)}{1+\exp\left(-\sqrt{\frac{2q^2}{\beta Mr}}\right)}}\right]
\right\}$\\
&\\
\hline
\end{tabular}
\caption{Energy distributions calculated by use of the energy-momentum complexes of Einstein, Landau-Lifshitz, Weinberg, and M\o ller}
\end{table}

We have also examined the limiting behavior of the energy in the cases $r\rightarrow \infty$, $q=0$, $\beta\rightarrow 0$, and $\beta\rightarrow \infty$. For $\beta\rightarrow 0$, the metric function $f(r)$ becomes $1-
2M/r$, while for $\beta\rightarrow\infty$, $f(r)$ becomes $1-(2M/r)[\exp(-q^2/2Mr)]$. The corresponding energies are presented in Table 2, where one can see that for $q=0$ as well as at infinity the Einstein and M\o ller prescriptions yield the same result which is also obtained for the classical Schwarzschild black hole solution, namely the ADM mass $M$. In the case of the Landau-Lifshitz and Weinberg prescriptions the energy equals the ADM mass $M$ at infinity $r\rightarrow \infty$, while in the chargeless case $q=0$ these two energy-momentum complexes give for energy the expression $M/(1-\frac{2M}{r})$. These results coincide with those obtained by Virbhadra's approach in Schwarzschild Cartesian coordinates \cite{Virb_1}. As it is pointed out at the end of Section 2, for $\beta \rightarrow 0$ we get the classical Schwarzschild black hole metric. Hence, the corresponding energies given in Table 2 for $\beta \rightarrow 0$ are those obtained in the four prescriptions for th!
 e Schwarzschild black hole. It is noteworthy that for $\beta \rightarrow \infty$ a factor of 2 appears in the exponential in all four prescriptions. In fact, the same dependence on this factor of 2 is obtained for the Einstein and M\o ller energies in \cite{Radinschi_AHEP}, where a different metric function $f(r)$ is considered.

\begin{table}[t!]
\centering
\begin{tabular}{|l|c|c|c|c|}
\hline
Prescription & Energy for & Energy for & Energy for &Energy for\\
 & $r\rightarrow \infty $ & $q=0$ & $\beta \rightarrow 0$ & $\beta
\rightarrow \infty $ \\ \hline\hline
&&&&\\
Einstein & $M$ & $M$ & $M$ & $M\exp \left( -\frac{%
q^{2}}{2\,M\,r}\right) $ \\ 

&&&&\\
\hline
&&&&\\
Landau-Lifshitz & $M$ & $\displaystyle{\frac{M}{1-\frac{2M}{r}}}$ & 
$\displaystyle{\frac{M}{1-\frac{2M}{r}}}$ & $\frac{M\exp(-\frac{q^{2}}{2\,M\,r}) }{1-\frac{%
2\,M}{r}\exp \left( -\frac{q^{2}}{2\,M\,r}\right)}$ \\ 
&&&&\\
\hline
&&&&\\
Weinberg & $M$ & $\displaystyle{\frac{M}{1-\frac{2M}{r}}}$ & $\displaystyle{\frac{M}{1-\frac{2M}{r}}}$ & $\frac{M\exp \left( -\frac{q^{2}}{2\,M\,r}\right) }{1-\frac{%
2\,M}{r}\exp \left( -\frac{q^{2}}{2\,M\,r}\right) }$ \\ 
&&&&\\
\hline
&&&&\\
M\o ller & $M$ & $M$ & $M$ & $[M-\frac{q^{2}}{2r}]\exp \left( -\frac{q^{2}}{2M\,r}\right) $
\\
&&&&\\ \hline
\end{tabular}%
\caption{Limiting cases for the energy}
\end{table}

\begin{figure}[h!]
\begin{center}
\includegraphics[width=84mm]{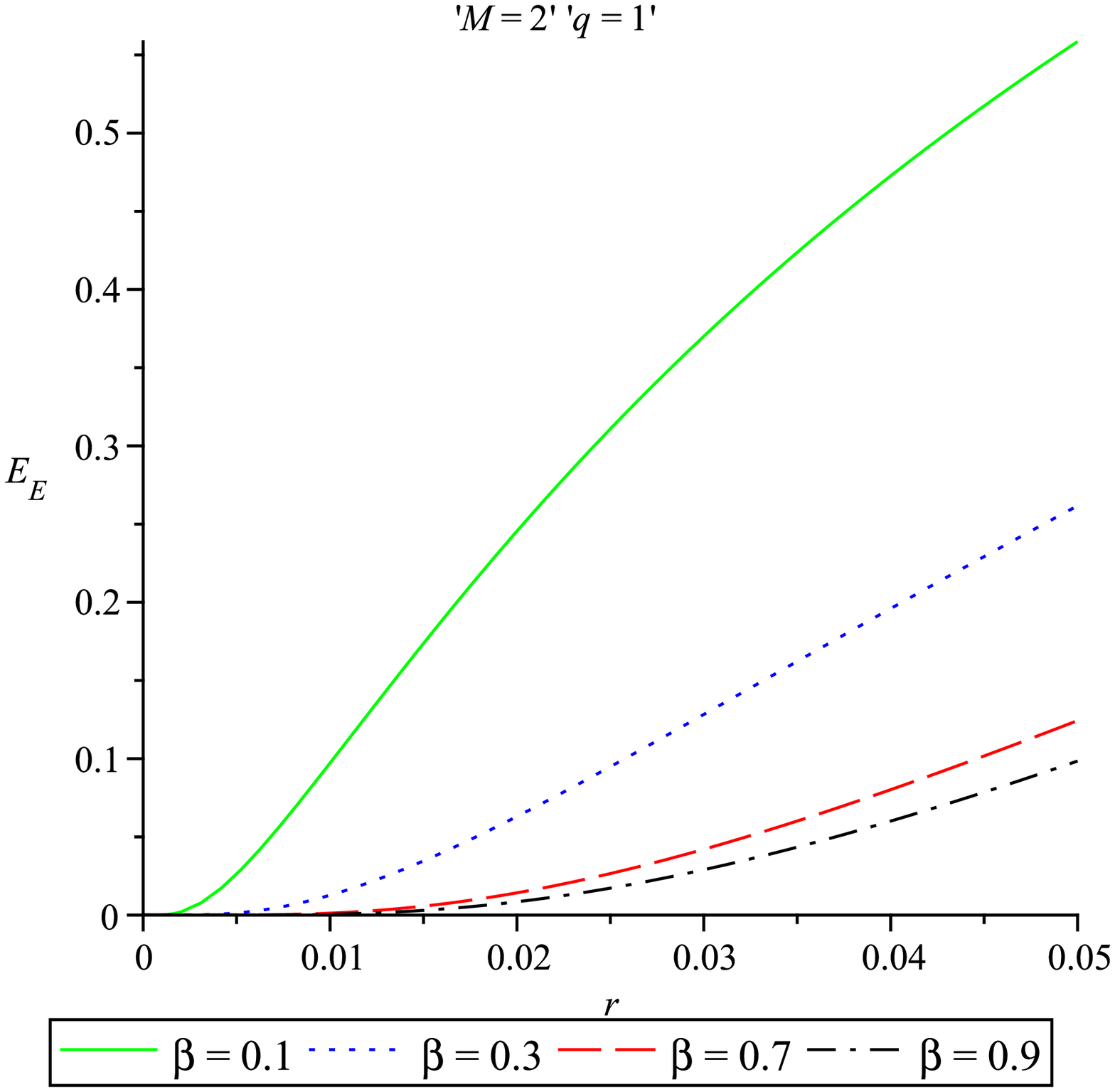}
\caption{Energy distribution near zero obtained in the Einstein prescription for different values of $\beta$.}
\label{fig1}
\end{center}
\end{figure}

\begin{figure}[h!]
\begin{center}
\includegraphics[width=84mm]{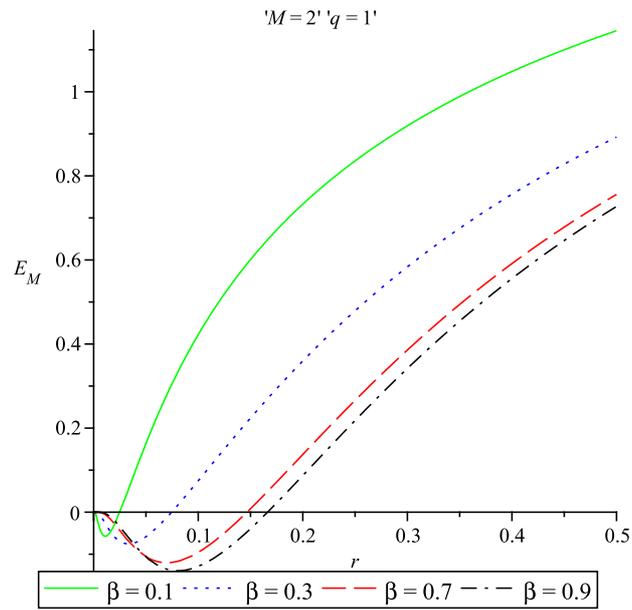}
\caption{Energy distribution near zero obtained in the M\o ller prescription for different values of $\beta$.}
\label{fig1}
\end{center}
\end{figure}

Of particular interest is the behavior of the energy near the origin, namely for $r\rightarrow 0$. In the case of the Einstein prescription, the energy tends to zero (see Figure 3). However, the energy obtained by the application of the Landau-Lifshitz, the Weinberg and the M\o ller energy-momentum complexes shows a rather pathological behavior. In the first two cases the energy is jumping between infinitely positive and negative values at some radial distances in the range $0<r<5$ for different values of the parameter $\beta$, while for significantly larger values of $r$ ($r>50$) the energy falls off rapidly to a constant value for various values of the parameter $\beta$. The energy calculated by the M\o ller prescription becomes clearly negative in the range $0<r<0.2$ for different values of the parameter $\beta$, while the position where the energy retains a positive value and then keeps increasing monotonically is shifted nearer to the origin as $\beta$ becomes smaller. !
 This strange behavior of the energy distribution may enhance the argumentation (see, e.g. \cite{Virb_1})  according to which the Einstein  energy-momentum complex is indeed a more reliable tool for the study of the gravitational energy-momentum localization as it yields physically meaningful results. Thus, it may gain a preference among the different pseudotensorial energy-momentum complexes.

\section*{Acknowledgements}

The authors thank the anonymous referees for their valuable comments and suggestions. Farook Rahaman is grateful to the Inter-University Centre for Astronomy
and Astrophysics (IUCAA), India, for providing Associateship Programme. Farook Rahaman and Sayeedul Islam are thankful to DST, Government of India for providing financial support
under the SERB and INSPIRE programme. 

\section*{Statement on conflict of interests}

The authors declare that there is no conflict of interest regarding the publication of this paper. Also, there is no conflict of interest regarding the received funding mentioned in Acknowledgement section. Also, there are no other possible conflicts of interests in the manuscript.

\end{document}